# Probing frustrated metallo-molecular spin Kondo lattice interfaces through anomalous Nernst effect

Servet Ozdemir,[1,*] Matthew Rogers,[1] Conor J. McCluskey,[2] Raymond G. McQuaid,[2] Zachariah Parkin,[1] Mannan Ali,[1] Gavin Burnell,[1] Joseph Barker,[1] B J Hickey,[1] and Oscar Cespedes[1,†]

[1]*School of Physics and Astronomy, University of Leeds, LS2 9JT, Leeds, UK*
[2]*Centre of Quantum Materials and Technologies, School of Mathematics and Physics, Queen's University Belfast, BT7 1NN, Belfast, UK*

Frustrated Kondo spin lattice (KSL) systems away from the antiferromagnetic (AFM) ground state have been found to display strange metal behaviour. A signature of strange metals in correlated systems is large Nernst response. Metallo-molecular interfaces of supramolecular lattices have been demonstrated as 2D KSL systems in STM studies. Here going beyond STM experiments we report a frustrated AFM state on molecular interfaces of Pt(111) and Pt(111)/Co films with around room temperature spin freezing transitions. Near these transitions we measure an anomalous Nernst coefficient of at least 3 $\mu$V/K.

Recent scanning tunnelling microscopy (STM) studies of Kondo resonance on metallo-molecular interfaces have led to observations of Ruderman-Kittel-Kasuya-Yosida (RKKY) coupling between molecular-spin sites, suggesting emergence of a metallo-molecular Kondo spin lattice accompanied by long range magnetic order [1–3]. Kondo spin lattices have led to the emergence of heavy fermion $f$ orbital systems that have been subject to prolonged interest [4, 5], with anti-ferromagnetism and superconductivity [6] emerging in these systems in the vicinity of quantum criticality [7]. The relative strength of the RKKY interaction between spin-lattice sites is mediated by the conduction electrons, and the single-ion Kondo interaction at each spin-lattice site dictates the nature of emerging phases, including the quantum phase transitions between them. Frustration has been understood to be another relevant parameter in Kondo spin lattice systems [8, 9], where a strange metal or a spin liquid state is expected in the so-called localised limit in the vicinity of the antiferromagnetic ground state[10].

Existing evidence in this limit in conventional systems has been obtained on bulk systems such as $Pr_2Ir_2O_7$ [11, 12], $YbRh_2Si_2$ [13, 14], CePdAl [15] and CeRhSn [16]. The phase transitions are observed at low temperatures, for example the freezing temperature $T_f$ for $Pr_2Ir_2O_7$ is 0.3 K [12] and the Néel temperature $T_N$ for CePdAl is $\approx$ 2.75 K [15]. Looking beyond existing 2D van der Waals systems [17–22], metallo-molecular thin film interfaces may offer much higher magnetic transition temperatures [23]. A better fundamental understanding of Kondo spin lattices [24] could also be gained through metal surface choices of different conduction electron bandwidths, but the observations so far have been limited to ultra-high vacuum STM experiments [1–3].

Here, we demonstrate magnetic frustration as well as a large anomalous Nernst effect on flat lying [1, 3, 25] metal (Cu, Co) and hydrogen ($H_2$) phthalocyanine (Pc) supramolecular lattice interfaces with textured Pt(111) and Pt(111)/Co(< 2 nm) films. The chamber pressure during the Pc layer sublimation was kept on the order of $10^{-10}$ mbar to enable a strong $\pi$-$d$ orbital hybridisation [26] during chemisorption [27] as well as the formation of a molecular lattice [1–3, 28]. We deposited a molecular layer thicker than 10 nm on top of Pt and Pt/Co films to ensure a continuous film of flat lying molecules at the interface. We capped the multilayer with a 15 nm thick metal layer to preserve the high-vacuum hybridised interface in ambient conditions.

The emergence of molecular Kondo spin lattices in STM experiments has been demonstrated for molecules deposited on Au(111) [1, 2] and Ag(111) [3] surfaces, which naturally makes hybridised organic molecular interfaces with Pt(111) a candidate system (see Fig. S1, Supplemental Material [29], for XRD peak). Fig. 1a shows the post zero field cooled (ZFC) and field cooled (FC) magnetisation curves as a function of temperature for a Pt(4.2 nm)/CuPc interface with an in-plane field of 20 mT. Fig. S2 [29] shows a similar result for a Pt/$H_2$Pc interface, ruling out a mechanism related to the presence of metal ions in the molecule. At room temperature, prior to ZFC cooling, the system is weakly magnetic, ($M_S$ = (49±1) emu/cm$^3$ for Pt/CuPc– see Supplemental Fig. S3, [29]), which we attribute to spin polarised charge transfer [30] from intrinsically paramagnetic Pt and enhanced exchange interaction [23, 31]. After zero field cooling, a close to zero-magnetisation (or close to zero magnetic susceptibility) state emerges close to base temperature, as it would be expected for an antiferromagnetic ground state. Magnetic susceptibility is peaking at $\approx$ 300 K as shown in Fig. 1a which could be interpreted as a Néel temperature. The rate of magnetisation onset peaks at 130 K as shown in the inset on Fig. 1a. During field cooling, the system is already switched to a ferrimagnetic state with a low temperature magnetisation $M_{FC} \approx$ 27 emu/cm$^3$ in a 20 mT applied field. This suggests that the measured ZFC state of zero susceptibil-ity is metastable, as it would be expected for a frustrated antiferromagnet, hence it would be more appropriate to refer to the peak value of magnetic susceptibility at 300 K as a spin freezing temperature. The observation of similar results in a Pt(111)/$H_2$Pc interface (Supplemental Material, Fig. S2, [29]) points at strongly $\pi$-$d$ hybridised, spin polarised hexagon-pentagon carbon units as shown in STS experiments [32]. This mechanism is depicted on Fig. 1c, with the spin lattice arrangement [1–3] shown in Fig. 1d. Using a model of RKKY coupling in 2D [33], and considering the spin-polarisation to be localised on the hexagon-pentagon carbon unit [32] we estimate [34] a nearest neighbour spin-lattice distance of 0.59 nm, and a next nearest neighbour distance of 0.83 nm on the hexagon-pentagon units (see Supplemental Material, [29], Fig. S4). It has been shown that the (111) surface states of Pt are Rashba spin split, and have Fermi wavevector ($k_F$) values of 1.4 nm$^{-1}$ and 1.9 nm$^{-1}$ [35]. In experiments studying graphene interfaces on Pt(111) surfaces, it has been shown that $\pi$ orbital hybridisation occurs with spin-polarised surface states at $k_F$ value of 1.4 nm$^{-1}$ [36]. Taking this value we calculate a nearest neighbour exchange parameter $J_1$ = -0.81$J_0$ (where $J_0$ is the ferromagnetic $\pi$-$d$ orbital exchange interaction constant) and a next nearest neighbour exchange parameter $J_2$ = -0.34$J_0$ (see [29], Supplemental Note 1). It is known from studies of square lattices that the $J_1/J_2$ ratio quantifies the degree of frustration in the system. Our value of 0.42 suggests strong frustration effects [37, 38], in agreement with the metastability suggested by the mag-netometry measurements in Fig. 1a.

* S.Ozdemir@leeds.ac.uk
† O.Cespedes@leeds.ac.uk

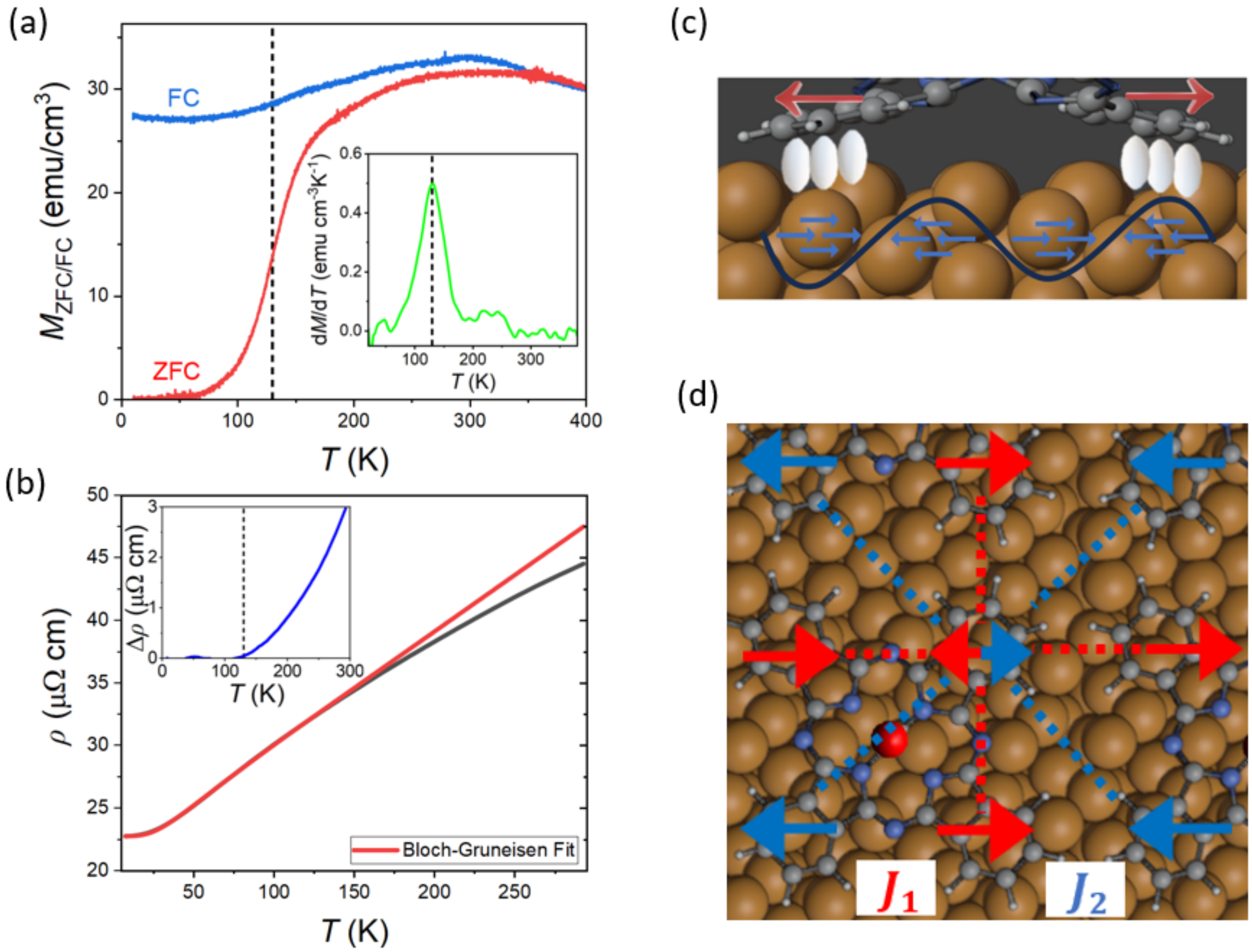

FIG. 1. Frustrated antiferromagnetism onset on (111) textured Pt/molecular interface with room temperature spin freezing point. (a) Zero field cooled - field cooled (ZFC-FC) magnetometry curves measured in a 20 mT in-plane field on a Pt(4.2 nm)/CuPc metallo-molecular interface, showing a broad spin freezing transition point peaking at $T_f \approx 300$ K . The derivative of the ZFC curve as a function of temperature is plotted on the inset and shows a peak at $T = 130$ K . b, Resistivity versus temperature measured on a strip of capped Pt(4.2 nm)/CuPc interface yielding a $T_D$=(163 ± 1) K with the red fit being the Bloch-Grüneisen fit that a Pt film is expected to follow. The deviation from the fit at high temperatures is plotted on the inset (blue curve), with the onset (dashed line) emerging at the $\mathrm{d}M/\mathrm{d}T$ peak at $T = 130$ K. c, Side view of hexagon-pentagon carbon units with $\pi$-orbitals hybridised to Pt(111) surface atoms, leading to spin polarised charge transfer and an RKKY-mediated interaction between them. d, RKKY coupled square lattice of spins formed on hexagon-pentagon units with the surface states of Pt mediating the interaction.

We investigated the Pt/molecular interface further using electrical transport measurements, Fig. 1b. Single-ion Kondo effects, typically observed as an increase in the low temperature resistivity [39] were not present. This absence is expected in the limit where RKKY interactions dominate over single-ion Kondo effects [10, 40]. The resistivity vs temperature of both a Pt/CuPc interface and the Pt reference samples can be described by the Bloch-Grüneisen law (see [29], Supplemental Information, Fig. S5 and Note 2), with a Debye temperature of $T_D = (163 \pm 1)$ K. Interestingly, there is a deviation from Bloch-Grüneisen behaviour starting at 130 K for the Pt/CuPc interface (inset dashed line). This onset seems correlated with a change in the magnetic susceptibility of the film (dashed line Fig. 1a). Chemisorbed CuPc molecules have been previously shown to be non-dynamic at low temperatures [41], so we ascribe the deviation from Bloch-Grüneisen law and the finite magnetisation onset at 130 K to a vibrational coupling between Pt and the hybridised molecules as the metal phonon modes are fully occupied approaching the Debye temperature. Hence, we attribute the lower than expected resistance in the Pt/molecule interface to enhanced parallel conduction across the molecules hybridised to metal, which becomes comparable to that of across the Pt metal channel and is thermally activated. An Arrhenius fit to the resistivity deviation from the Bloch-Grüneisen fit yields a thermal activation energy of (76 ±1) meV (inset on Fig. 1b and [29], Fig. S6). This number is in a good agreement with the 70-120 meV energy gap previously measured in organic charge transfer interfaces [42].

A question emerges as to whether a frustrated antiferromagnetic state could also emerge at a Pt(111)/Co(1.7 nm) – organic molecule interface. These systems show a perpendicular magnetic anisotropy (PMA) at room temperature [43] and x-ray magnetic dichroism sum rule analysis, post field cooling, has indicated a scaling relation that we attribute to the frustrated ferromagnetic Kondo spin lattice [43]. The ZFC-FC magnetometry curves measured for a Pt/Co(1.7 nm)/CuPc interface in a planar magnetic field of 20 mT are shown on the upper panel of Fig. 2. A transition with $T_f = 237$ K is apparent, with very low magnetisation measured below $T_f$ in the ZFC curve, suggesting an antiferromagnetic state, as in the case of Pt/molecule interfaces. The surface states on Co films have been studied extensively [44, 45], and these states mediate the RKKY interaction between hexagon-pentagon units. Similarly to Pt/CuPc, a deviation from the Bloch-Grüneisen law is measured for Pt/Co/molecular interfaces (see [29], Supplemental Material, Fig. S7), suggesting that vibrational modes in the metal suppress the molecular lattice spin interactions above $T_f$. The ZFC spin-freezing peak is much sharper for Pt/Co/molecule than in Pt/molecule interfaces, which we attribute to the emergence of a perpendicular spin orientation on the Pt/Co interface above $T_f$ [43]. The bottom panel on Fig. 2 shows the in-plane magnetic remanence measured on the same sample at zero field after a 2 T field cooling, where we can observe that the system switched to an in-plane ferromagnetic state for $T \leq T_f$ (see [29], Fig. S8 for magnetic hysteresis above and below $T_f$).

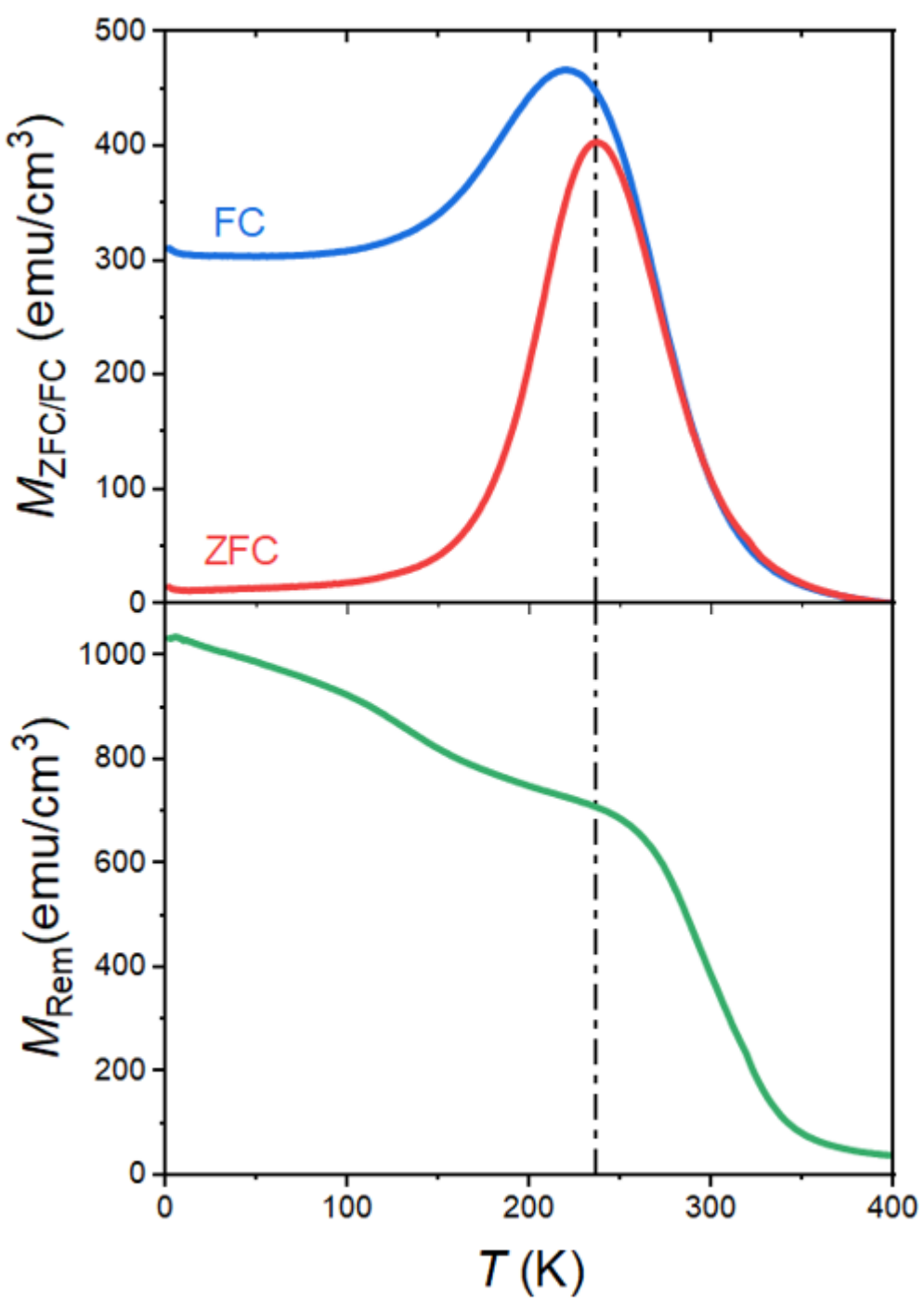

FIG. 2. Planar frustrated antiferromagnetism onset in perpendicular magnetic anisotropy Pt/Co bilayers with a copper phthalocyanine interface. A Pt/Co(1.7 nm)/CuPc interface has an out-of-plane easy axis at high temperatures with negligible in-plane remanence after saturation. Zero field cooled - field cooled in-plane magnetometry curves (upper graph – 20 mT) and in-plane remanence measured at zero field after field cooling in a 2 T field (lower graph) yield a freezing temperature $T_f \approx 237$ K that coincides with the in-plane magnetisation onset and higher remanence state attributed to the formation of the Kondo spin lattice.

Near quantum phase transitions away from the frustrated antiferromagnetic ground state, non Fermi-liquid behaviour is expected [8, 10] and it has been found in the form of emergence of a strange metal state [46]. A characteristic signature of a strange metal state that is widely observed in cuprates as well Kondo spin lattice systems is a linear temperature dependent resistivity [40]. However, given the interfacial nature of our metallo-molecular Kondo spin lattice system, we have significant paralell conduction from Pt/Co layers, and the molecular layers (above 130K), hence leading to the system obeying Bloch-Gruneisen law and deviating from it above 130K. An alternative powerful probe of emergence of a strange metal state is a manifestation of a sizeable (anomalous) Nernst effect [21]. Thus, to probe the phase transition further, we studied the Joule heating-induced Nernst effect, here superimposed on anisotropic magnetoresistance (AMR) [47–49] measurements for Pt/Co/molecular systems as shown in Figure 3a. The emergence of an unusual sinusoidal term [50] in the AMR curves at high current densities is due to a Nernst effect where the thermal gradient perpendicular to the sample plane is induced by Joule heating (Fig. 3a, 3b and [29], Supplemental Note 3)[47–50]. To clarify the nature of this asymmetry in the magnetoresistance with the magnetic field polarity, its magnetic field dependence was studied, see Figure 3c. The asymmetric term was found to dominate approaching zero magnetic field in the absence of AMR, hence confirming the anomalous nature of the Nernst effect (Fig. 3d).

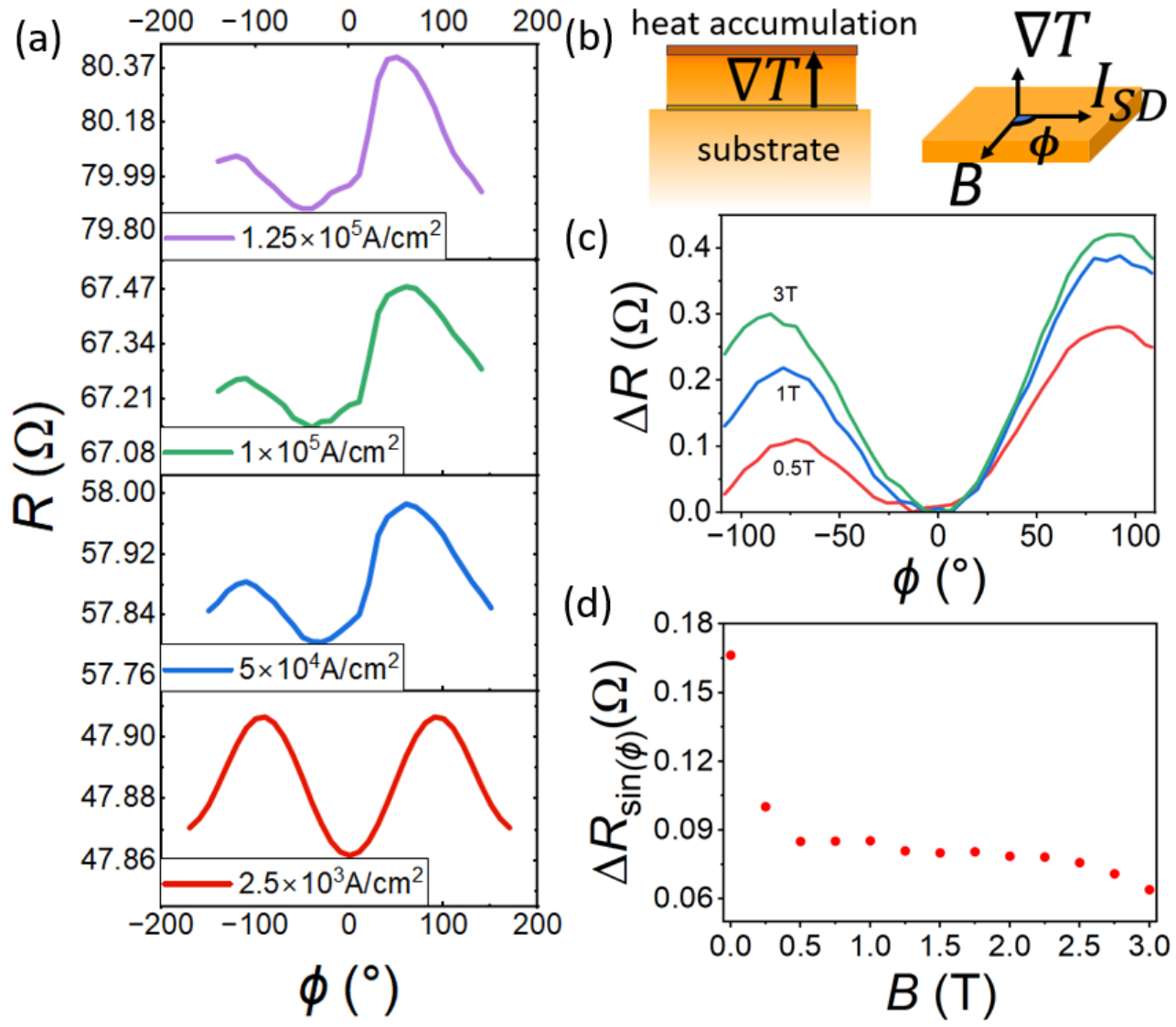

FIG. 3. Joule heating induced Nernst effect superimposed to anisotropic magnetoresistance on a Pt/Co/molecular interface. (a) DC measurements (averaged opposite polarities of current) of the planar anisotropic magnetoresistance (AMR) at various current densities as a function of planar angle ($\phi$) between current and magnetic field, where $B = 3$ T, on Pt/Co(1.7 nm)/CoPc metallo-molecular structure at $T = 150$ K. (b) Schematically illustrated i) Joule heating-induced perpendicular to sample plane temperature gradient generated due to heat accumulation and ii) source-drain current, temperature gradient, and magnetic field. (c) AMR curve measured with current density of $6.5 \times 10^5$ A/cm$^2$ at varying magnetic fields on a Pt/Co(1.4 nm)/$H_2$Pc metallo-molecular interface at $T = 250$ K. (d) Magnetic field dependence of fitting extracted amplitude of Nernst effect induced $\sin(\phi)$ term on a $H_2$Pc metallo-molecular interface at $T = 250$ K.

To clarify the role of the phase transition, we measure the ANE at three different temperatures (all post field cooling at 3 T): before, during and after the phase transition, as shown in Fig. 4a. This was carried out with a current density of $3.7 \times 10^5$ A/cm$^2$, which keeps the temperature increase of the sample below 20 K in Pt/Co(1.4nm)/$H_2$Pc interfaces (see, [29], Supplemental Information, Fig. S9 for temperature interpolation). The largest change in normalised resistance is observed at a cryostat temperature of 250 K, in the vicinity of the easy axis switching for these samples (see the inset remanence curve on Figure 4a) suggests that the ANE is linked to the phase transition, as it would be expected for a strange metal state emerging in the vicinity of a Fermi surface transition at the metallo-molecular Kondo spin lattice interface [21, 51, 52]. The evolution of the anomalous Nernst voltage at different current densities measured at a cryostat temperature of 250 K is plotted on Figure 4b. To estimate the anomalous Nernst coefficient $S_{ANE}$, we performed finite element modelling (FEM) of the Joule heating (see [29], Methods, Figs. S10 and S11). Using the temperature drop across the organic layer to obtain varying values of $\nabla T_z$, we obtained the maximum thermal gradient and therefore lowest $S_{ANE}$ estimates for different current densities as shown in Figure 4c. The anomalous Nernst electric field is calculated using the expression $E_{ANE}=S_{ANE}\nabla T_z$. At the peak $E_{ANE}$ value of 7 V/m, calculated using a FEM estimated $\nabla T_z$ value of $2.3 \times 10^6$ K/m, we find $S_{ANE}$ to be 3.0 $\mu$V/K. This is considerably higher than e.g. the values for chiral Kagome antiferromagnets $Mn_3Ge$ [53] and $Mn_3Sn$ [54], and comparable to systems such as the canted antiferromagnet $YbMnBi_2$ [55] and the iron based binary ferromagnet $Fe_3Ga$ [56]. Previously, an approximately zero value of anomalous Nernst effect ($E_{ANE} \approx 0$ V/m) had been measured on in-plane magnetised Pt/Co films in experiments utilising a perpendicular heat gradient [47]. Contrary to our experiments, an ANE attributed to magnetisation was measured utilising a planar heat gradient and 10 repeats of a perpendicularly magnetised Pt/Co layers [57]. Here, we report a value of $S_{ANE}$ that is approximately a factor 3 larger at a remanent field of 3 mT. Furthermore, the common orientation of the magnetisation and thermal gradient [43], at a current density of $6.5 \times 10^5$ A/cm$^2$, where maximum Nernst coefficient is measured, imply that this effect is linked to large entropy of charge carriers around the quasiparticle breakdown point or namely a strange metal state emergence during the phase transition at the metallo-molecular interface rather than presence of a finite magnetisation.

In conclusion, we have demonstrated the emergence of frustrated spin order in metallo-molecular interfaces grown on highly textured Pt(111) layers with and without an intrinsically magnetic sub 2 nm Co layer. The interfaces, grown in ultra-high vacuum ($10^{-10}$ mbar), show two-dimensional Kondo spin lattice effects previously observed in scanning tunnelling microscopy experiments. Single-ion Kondo effect [39] signatures are absent from transport measurements, as expected in the localised limit of dominating RKKY interaction between spin sites. In comparison with previous studies, the unconventional systems reported here show very high spin freezing temperatures ranging from 240 K to 300 K. A giant Joule-heating induced ANE is observed near spin-freezing transition, which we attribute to a strange metal state onset in the vicinity of the Fermi surface transition [28]. The enhanced thermoelectric properties lead to a $S_{ANE}$ approaching 3 $\mu$V/K around room temperature with measured with current densities of the order of $10^5$ A/cm$^2$. This large coefficient, comparable with colossal ANE systems [55, 56] and combined with the possibility of tuning the transitions around room temperature, bring Kondo spin lattices closer to applications.

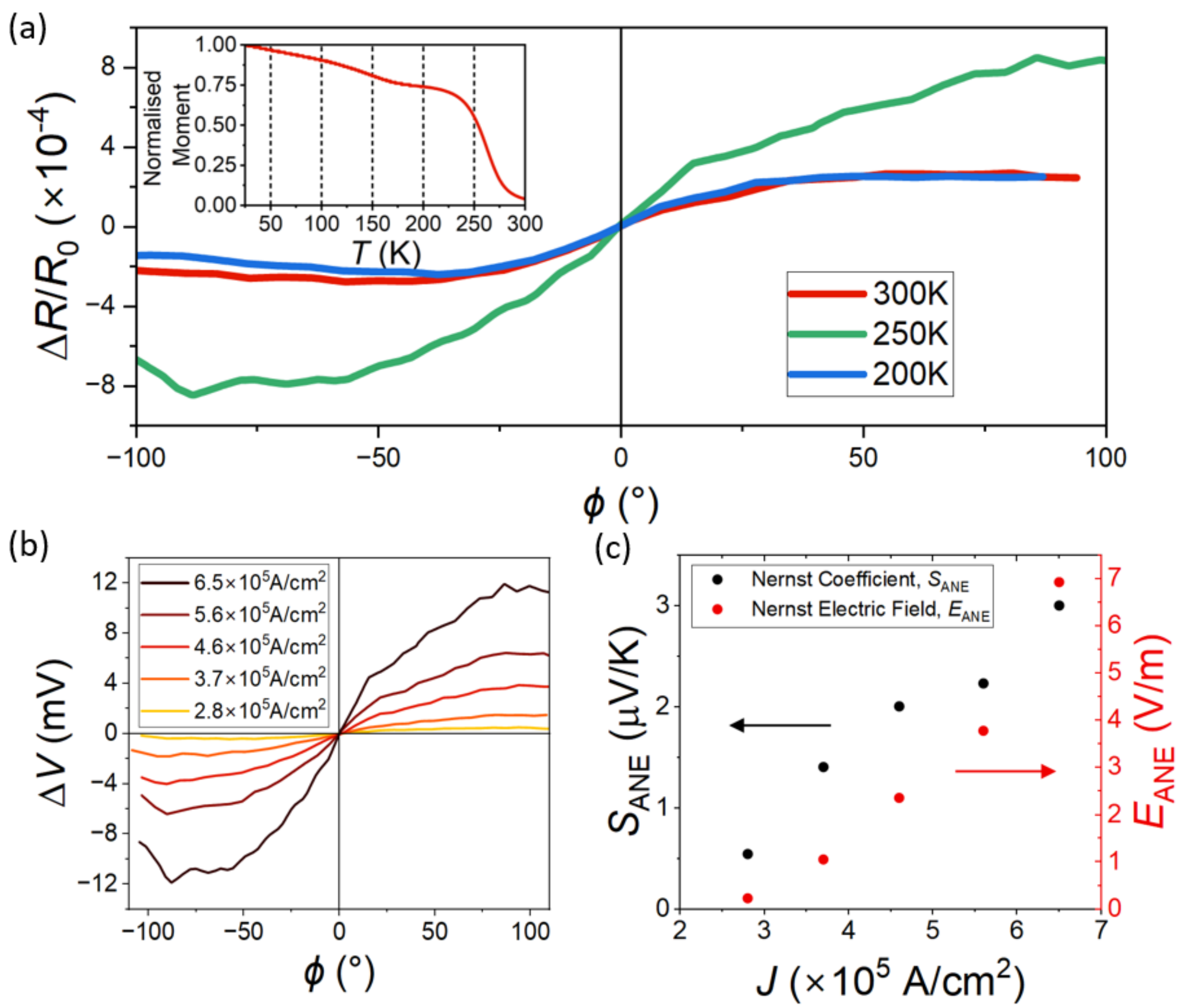

FIG. 4. Anomalous Nernst effect measured on Pt/Co/molecular interfaces in the vicinity of an easy axis transition (a) Normalised change in resistance as a function of angle ($\phi$) between remanent planar magnetic field and current due to anomalous Nernst effect (ANE) measured at three different temperatures around the easy axis switching region post field cooling at 3 T (see inset for the in-plane remanence curve for the corresponding Pt/Co(1.4 nm)/$H_2$Pc film) with a current density of 3.7 $\times$ $10^5$ A/cm$^2$. (b) Measured change in DC longitudinal voltage at $T$ = 250 K as a function of planar angle ($\phi$) between remanent magnetic field and current across a device at different Joule heating current densities. (c) Calculated anomalous Nernst coefficient ($S_{ANE}$) and measured anomalous Nernst electric field ($E_{ANE}$) at $T$ = 250 K as a function current density, with maximum ($S_{ANE}$) being 3 $\mu$V/K and maximum ($E_{ANE}$) measured as 7 V/m .

We acknowledge useful discussions with Valentin Irkhin and Ahmet Yagmur. We thank the Engineering and Physical Sciences Research Council in the UK for financial support via the grants EP/S030263/1 and EP/X027074/1. We also acknowledge the support of the EC project INTERFAST (H2020-FET-OPEN-965046). S.O. acknowledges the support of the Henry Royce Institute for Advanced Materials for enabling access to the Royce Deposition System facilities at the University of Leeds (EPSRC Grant EP/P022464).

# Methods

## A. Thin film growth and characterisation

Thin-film structures were grown on 0.65 $\mathrm{mm}$ thick c-plane sapphire films. The textured Pt(111) films (≈4 $\mathrm{nm}$) (See Fig. S1 for the XRD peak) were grown at 500 °C with e-beam evaporation at a growth rate of ≈0.1 Å/s. For samples also possessing the Co layer, the substrate was than cooled down to room temperature and the Co-layer was grown also with e-beam evaporation technique at a rate of ≈0.1 Å/s. Within the same chamber, organic molecule layers were sublimed onto the Co or Pt surface at a pressure $\approx 5 \times 10^{-10}$ mbar with a rate of ≈0.3 Å/s also at room temperature and until a thickness of $20$ nm was monitored through quartz monitor. Cap layer was magnetron sputtered on top of the thin film structure, with the material used as a cap being Cu for Pt/Co/molecule interfaces and Nb for Pt/molecule interfaces with a thickness of $\approx 15$ nm. Films were then structurally characterised using x-ray reflectivity (see Fig. S12) and x-ray diffraction. Molecular layers were characterised using Raman spectroscopy (see Fig. S13). Transmission electron microscopy characterisation of similar structures can be found in elsewhere [1,2].

## B. Magnetometry

Magnetisation measurements were carried out using SQUID-magnetometer (MPMS Q.D.) which offers a resolution over $10^{-8}$ emu accompanied by temperature control.

## C. Electronic transport

4-probe resistance measurements were carried out in continuous flow cryostats on devices of narrow cleaved strips of films, with a Keithley 6621 DC and AC current source with a Keithley nanovoltmeter being used for Joule heating induced anomalous Nernst effect measurements.

## D. Finite element modelling

Finite element modelling (FEM) was performed using the Joule heating Multiphysics interface on the commercially available COMSOL Multiphysics package. Electrically insulating boundary conditions were assumed for outer edges of the sample, except at the electrically contacted edges of the heterostructure, where a fixed current terminal and ground were supplied. The relative electrical resistivity of the oxidised copper layer, in comparison to the other metallic layers, was increased from the nominal copper resistivity until the maximum temperature gradient across the organic film was achieved. Then, the resistivity of all current carrying layers was increased to match the average resistivity of the experimental sample. For computational efficiency, the thermal conductivity of the substrate was increased so that the equivalent thermal resistance of the sapphire bulk could be obtained with only $500$ nm modelled thickness. An additional thermally resistive layer was added, below the sapphire substrate, which accounted for the any other thermal resistances present between the base of the substrate and a fixed temperature point. The base edge of this layer was held fixed at $250$ K, and its thermal resistance was modified until the peak temperature of the modelled system matched that of the experimental sample (as informed by resistance versus temperature measurements). All other outer layers had thermally insulating boundary conditions.

## Supplemental Figures

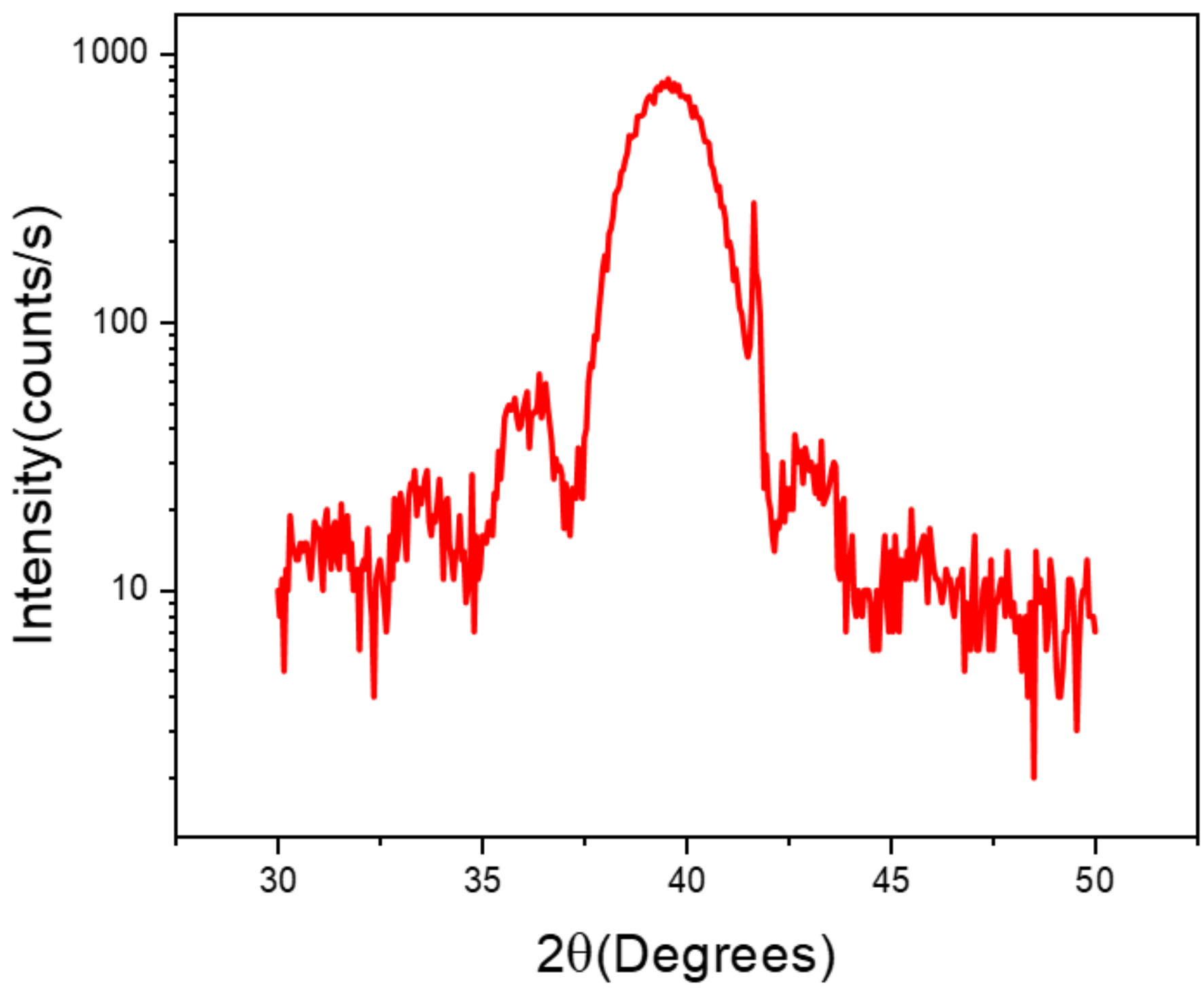

**FIG. S1. Pt film with (111) textured surface.** Pt(111) x-ray diffraction peak accompanied by Pendellösung fringes showing highly textured surface of the Pt film.

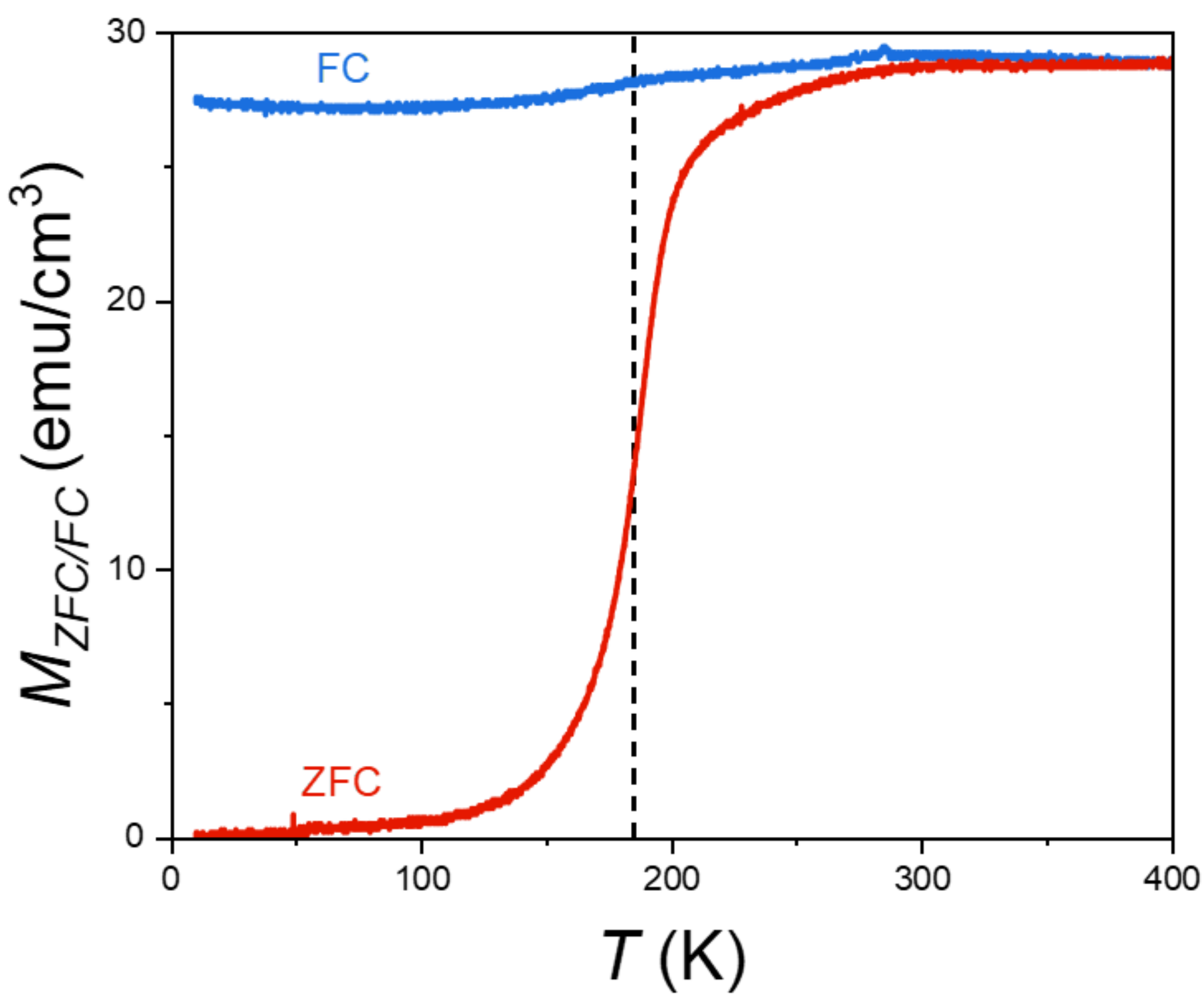

**FIG. S2. Magnetic frustration measured on a Pt/$H_2$Pc interface.** Zero field cooled (red) - field cooled (blue) magnetometry curves measured at 20 mT planar field on Pt(4.2 nm)/$H_2$Pc metallo-molecular interface, showing a broad spin freezing transition point peaking around 300 K.

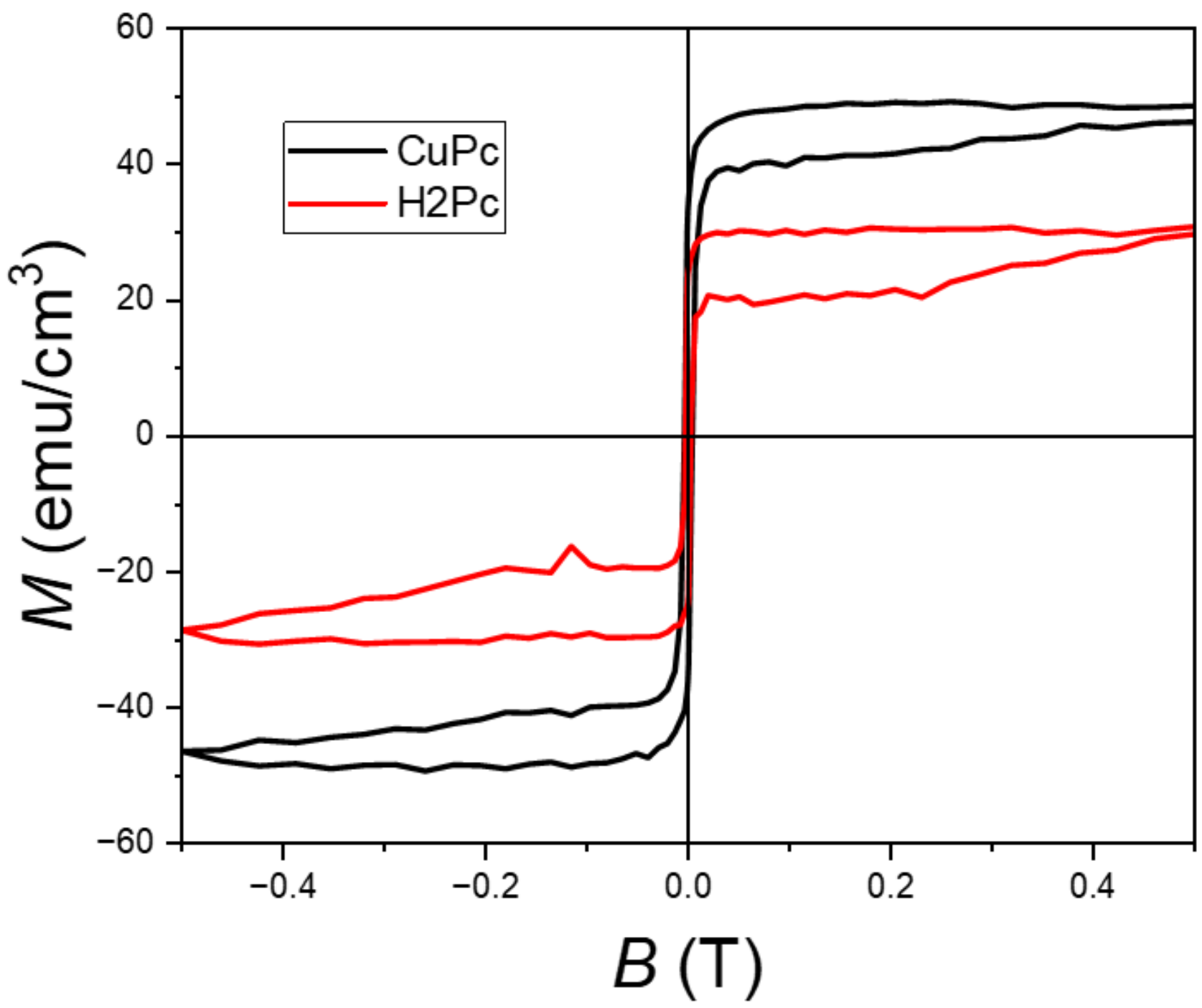

**FIG. S3. Room temperature magnetism emergence in Pt($4.2$ nm)/Pc interfaces.** Magnetisation curves measured for Pt(4.2 nm)/CuPc (black) and and Pt(4.2 nm)/$H_2$Pc (red) interfaces at 300 K, yielding respective saturation magnetisation values of (49±1) emu/cm$^3$ and (30±1) emu/cm$^3$.

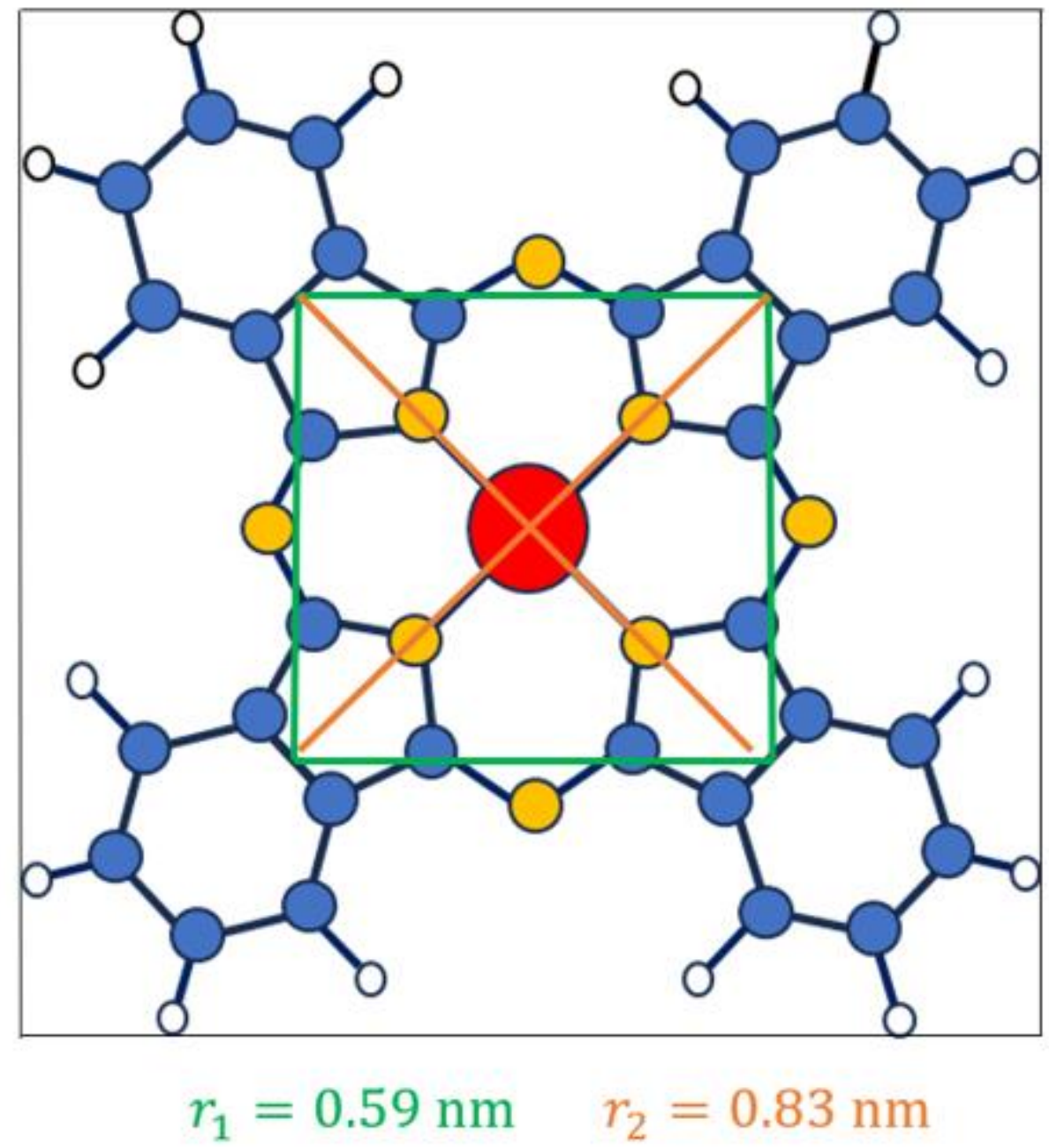

**FIG. S4. Geometry of a CuPc molecule.** Distances between centres of hexagon-pentagon units calculated using interatomic distances estimated for CuPc, leading to corresponding nearest neighbour, $r_1$ and next nearest neighbour $r_2$ distances on the spin lattice.

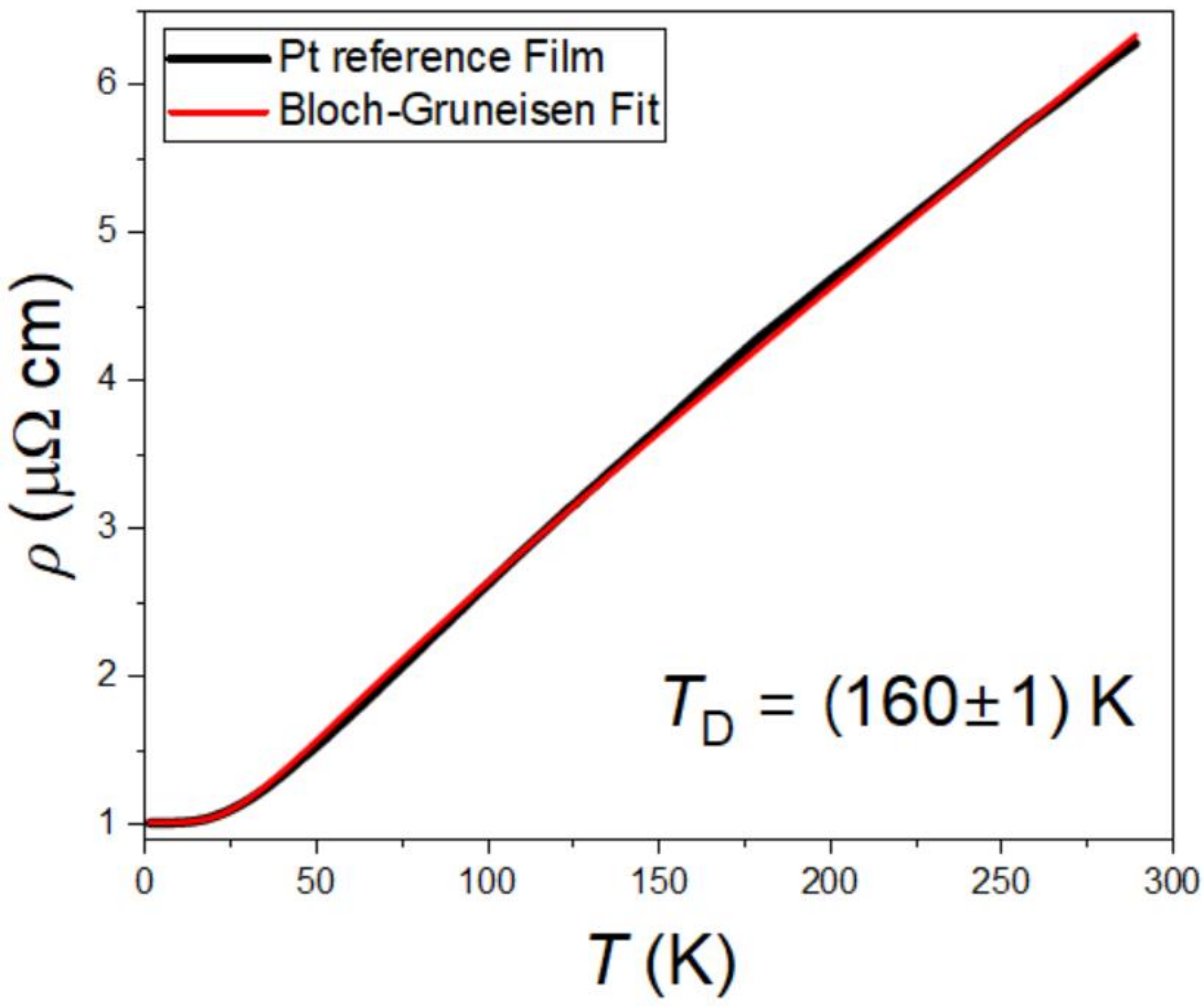

**FIG. S5. Temperature dependent resistivity and Bloch-Grüneisen fit of a Pt film.** The resistivity was found to be described by the Bloch-Grüneisen fit (Supplementary Note 1) as expected for a bare Pt film, with Debye temperature, $T_D$ obtained being $(160 \pm 1)$ K.

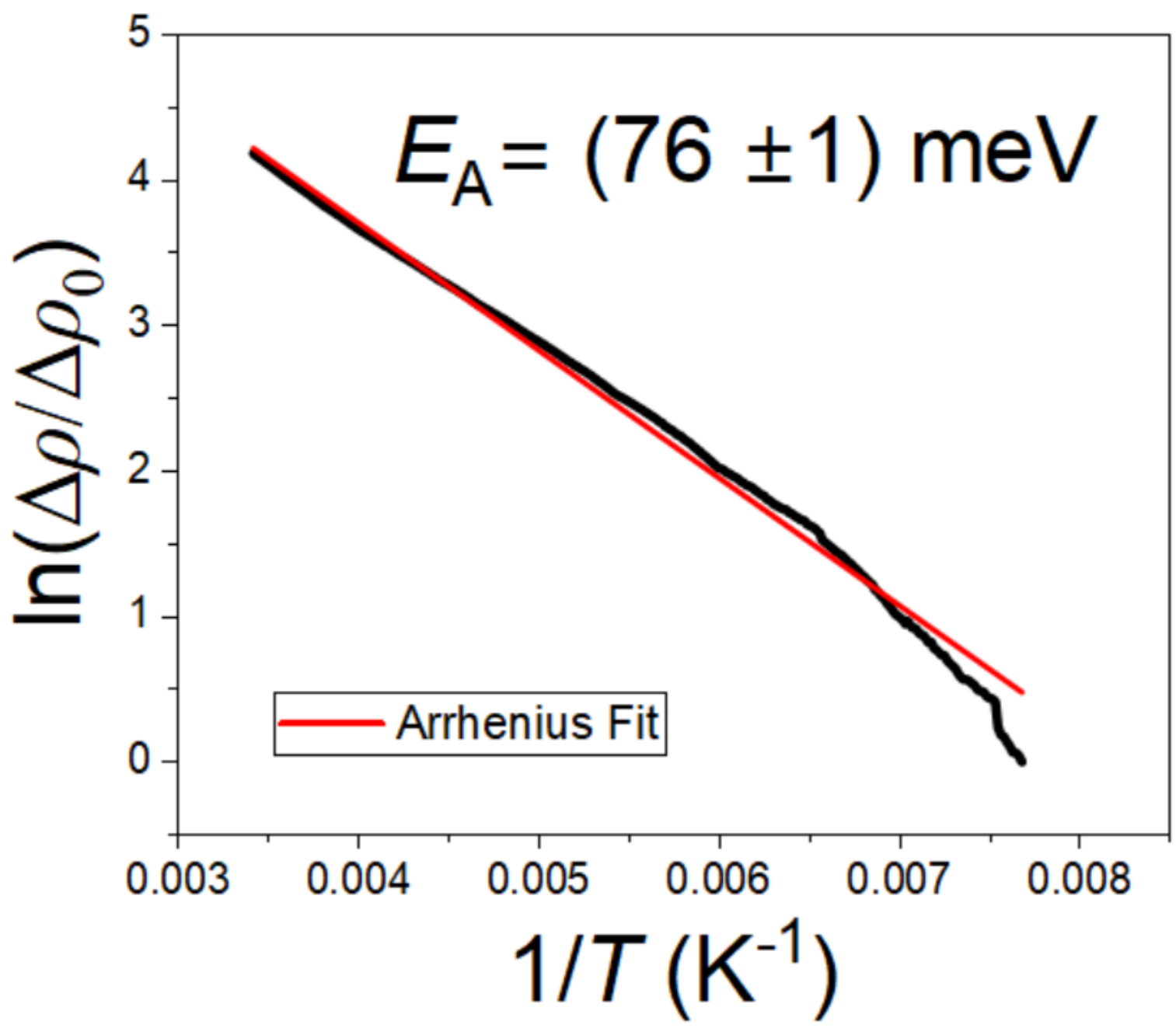

**FIG. S6. Thermally activated deviation from Bloch-Grüneisen fit on a Pt/molecular interface**. Arrhenius plot of resistivity difference corresponding to deviation from a Bloch-Grüneisen fit of temperature dependent resistance of Pt/CuPc interface, yielding a thermal activation gap of $(76 \pm 1)$ meV. The equation $\Delta\rho = \Delta\rho_0 e^{-E_A/k_B T}$ was used for Arrhenius fitting.

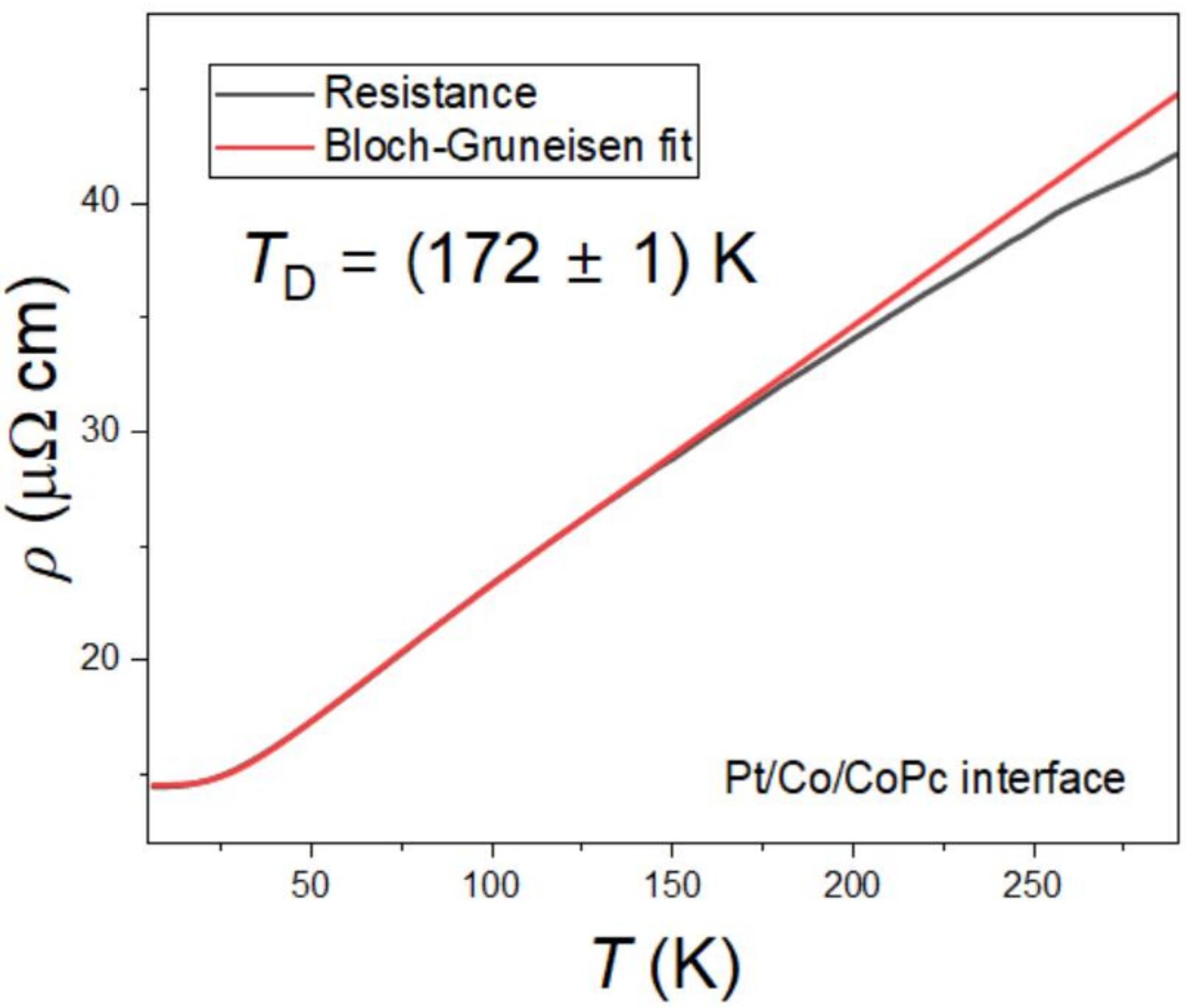

**FIG. S7. Temperature dependent resistivity on a Pt/Co(1.7nm)/CoPc interface.** The deviation from the Bloch-Grüneisen fit (Supplementary Note 1) is evident at higher temperatures, where the Debye temperature, $T_D$ , is obtained to be $(172 \pm 1)$ K.

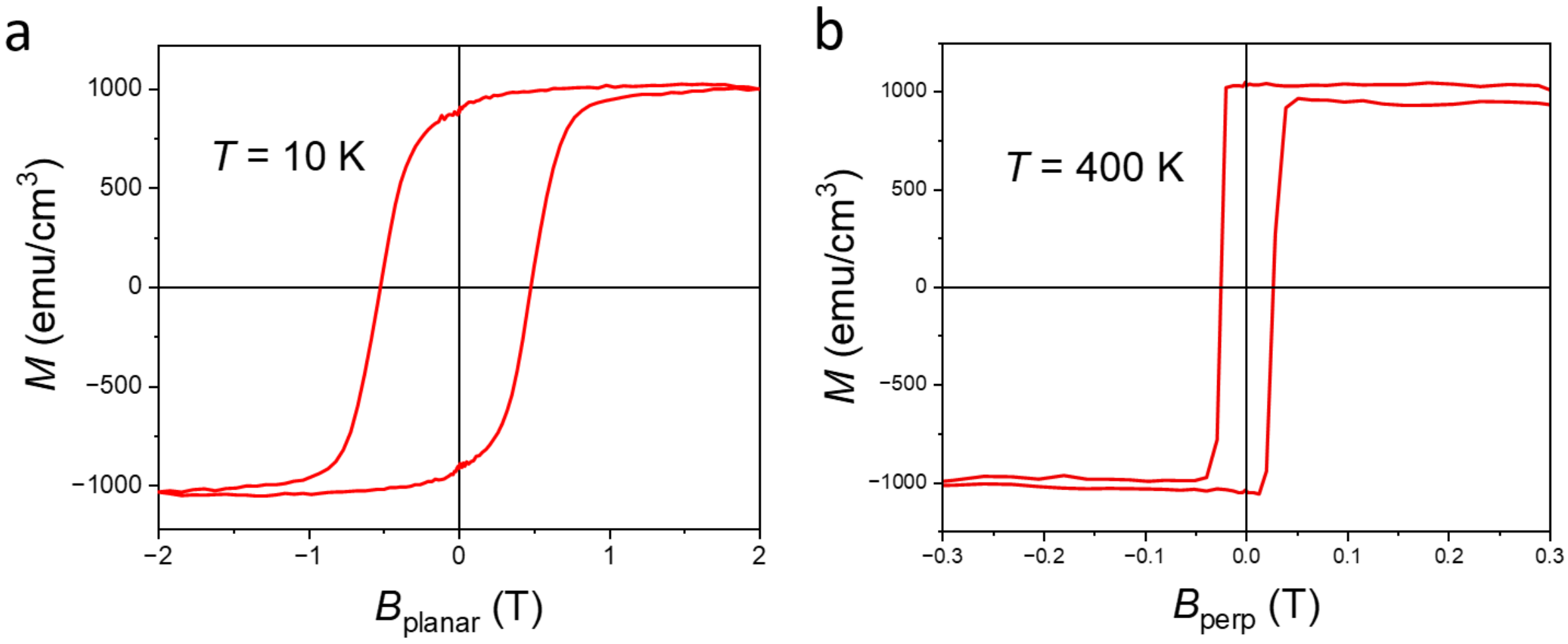

**FIG. S8. Hysteresis loops with planar and perpendicular magnetic fields measured on capped Pt/Co(1.7nm)/CuPc structure. a,** Planar hysteresis curve measured at *T* = 10 K where easy axis is in plane. **b,** Perpendicular magnetic field hysteresis curve measured at *T* = 400 K where easy axis is perpendicular to sample plane.

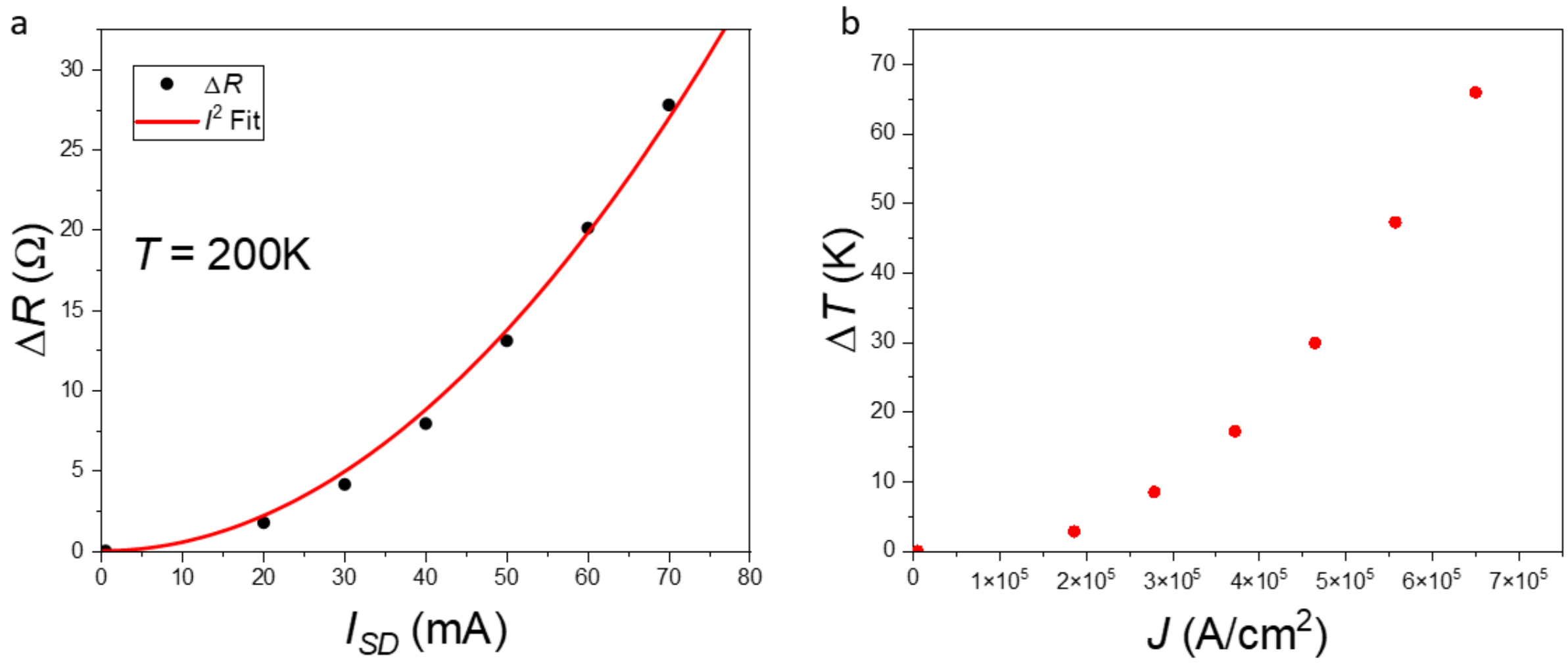

**FIG. S9. Joule heating induced temperature change on Pt(4.2 nm)/Co(1.4 nm)/$H_2$Pc device. a,** Measured resistance change as a function of current density at a cryostat temperature of 200 K. **b,** Cooling curve interpolated temperature changes corresponding to different current densities, suggesting a maximum temperature change of 66 K at a current density of $6.5\times10^5$ A/cm$^{-2}$.

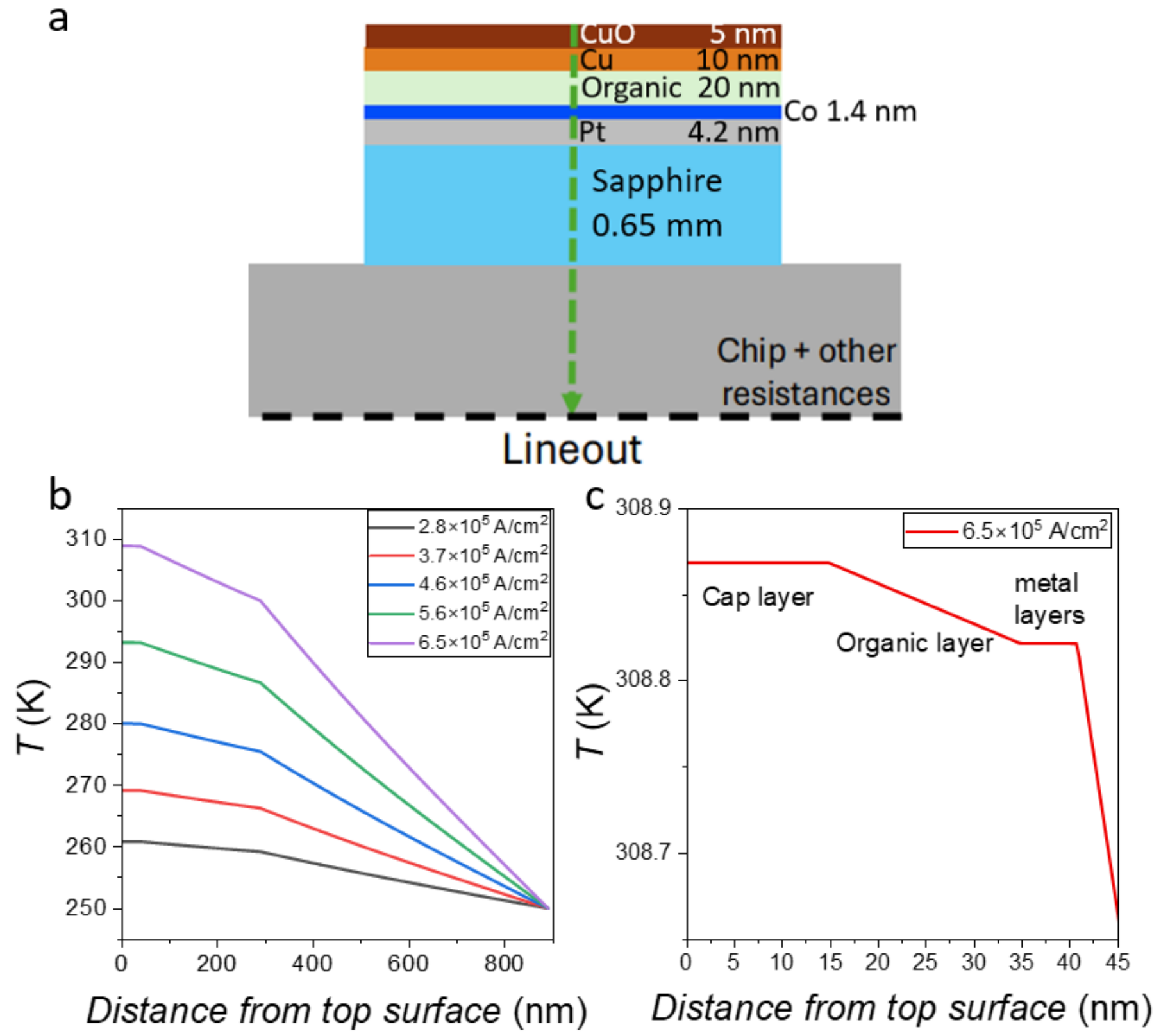

**FIG. S10. Finite element modelling of Joule heating of metallo-molecular structure. a**, Schematic of the structure for which finite element modelling was carried out. **b**, Temperature profile across the green line depicted in **a**, at various currents showing peak temperature accumulating at CuO layer to be 308 K for a current density of $6.5\times10^5\ \mathrm{A/cm^{-2}}$ where the base temperature of the system is 250 K. **c**, Zoomed in temperature profile across the metallo-molecular heterostructure for a current density of $6.5\times10^5\ \mathrm{A/cm^{-2}}$, showing temperature drop across the organic layer.

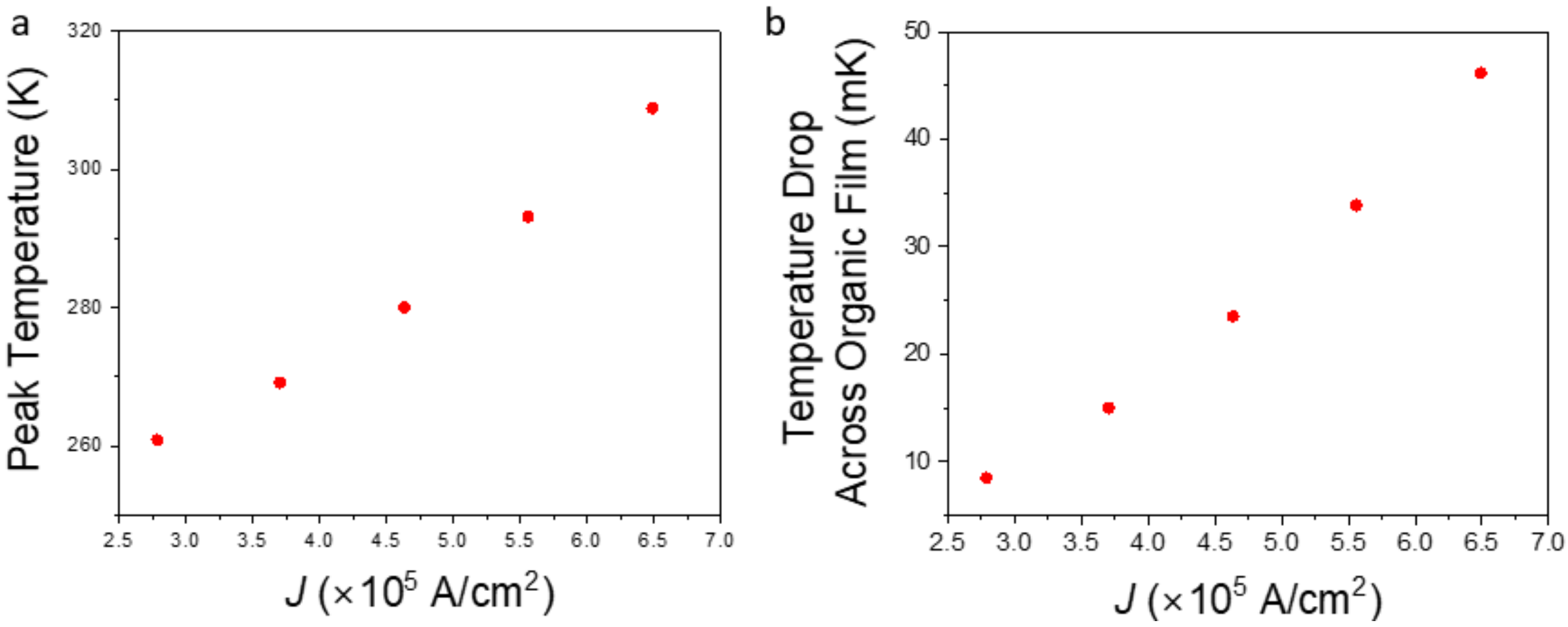

**FIG. S11. Finite element modelling results at different Joule heating current densities. a,** Peak temperature accumulating at CuO layer at different current densities that are studied experimentally. **b**, Temperature drop across the organic layer at different current densities at which experiments were carried out.

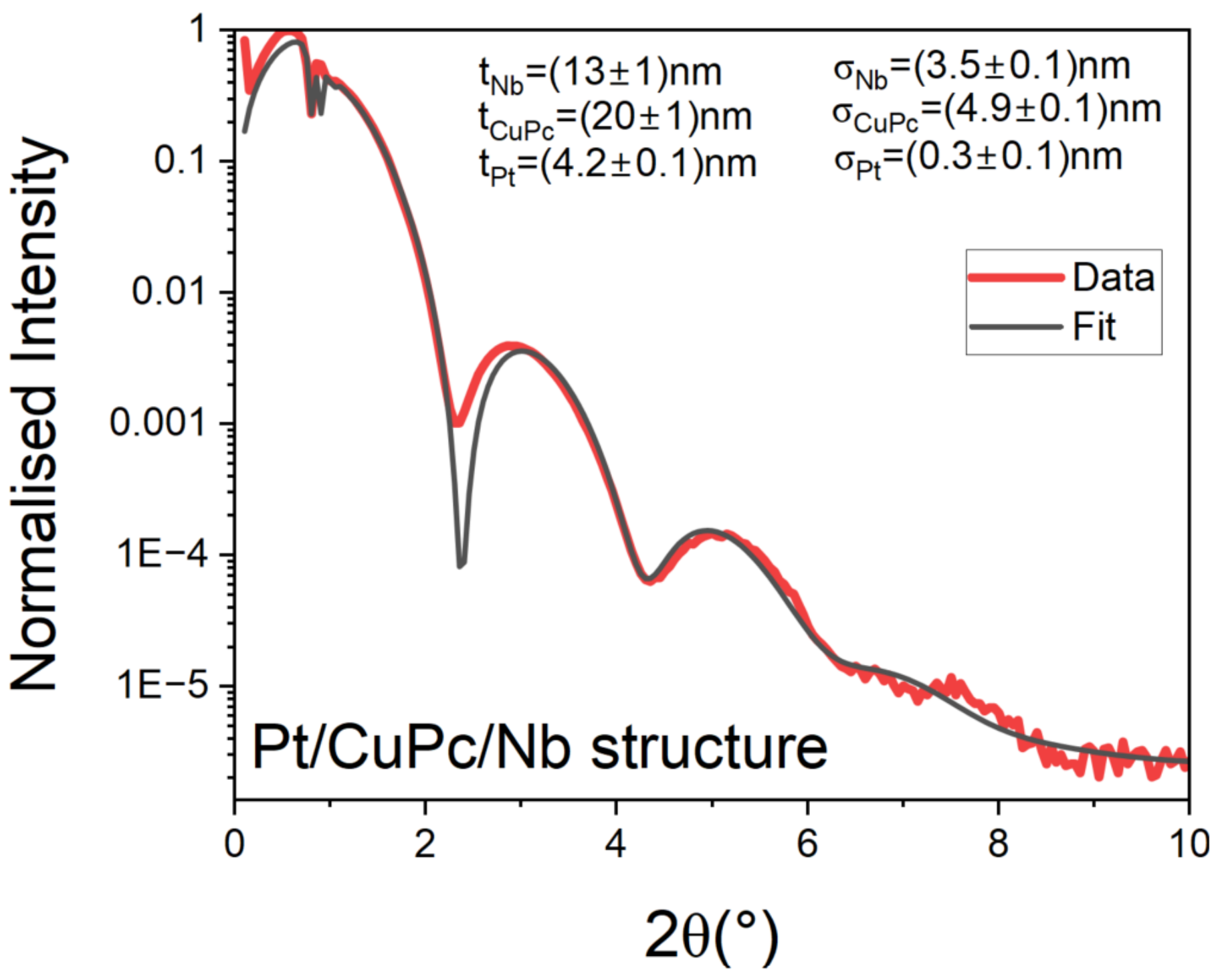

**FIG. S12. X-ray reflectivity of Nb capped Pt/CuPc film with obtained thickness and roughness given on the top right.**

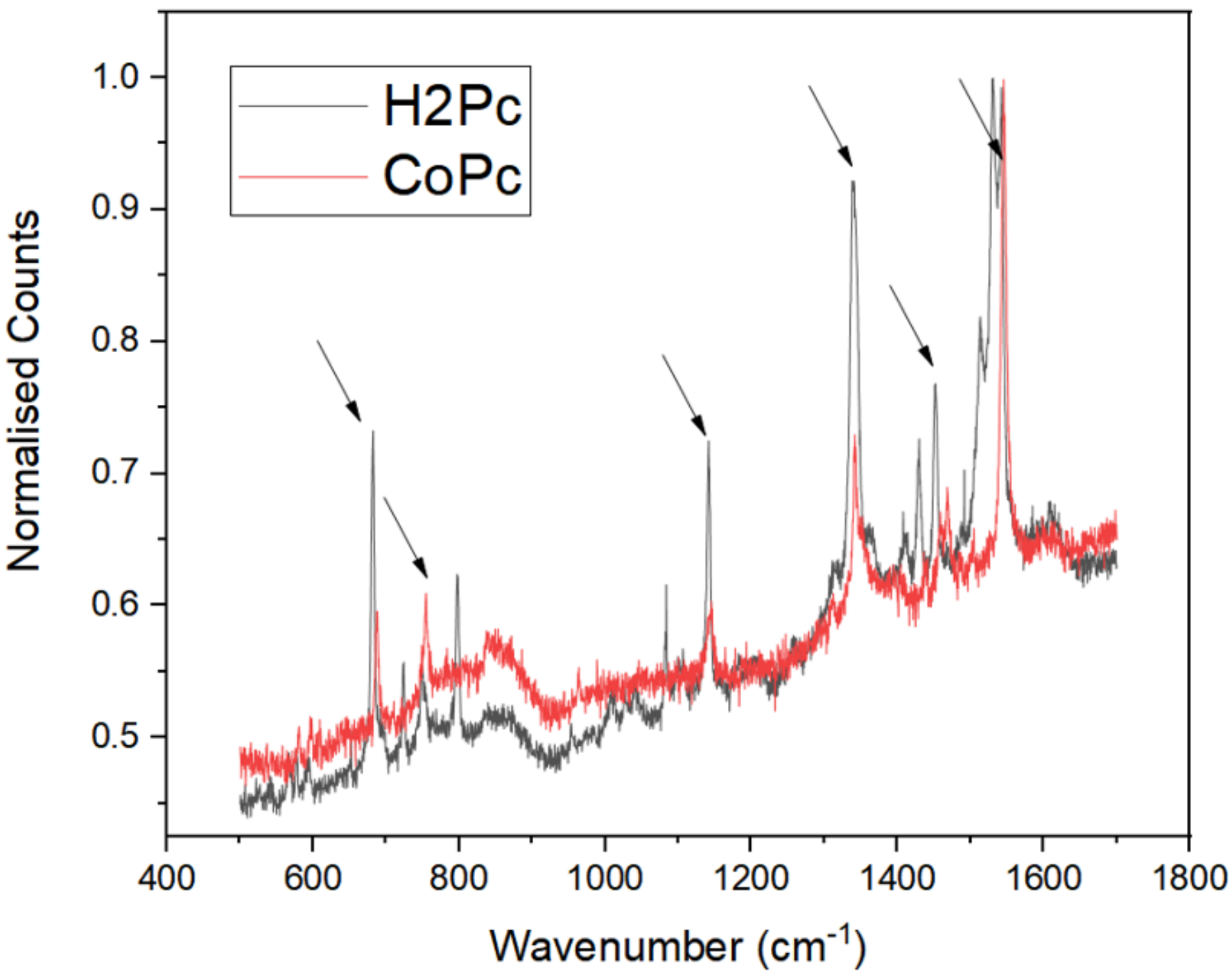

**FIG. S13. Room temperature Raman spectrum of a capped, Pt(4.2nm)/Co(1.7nm)/CoPc and Pt(4.2nm)/Co(1.7nm)/$H_2$Pc heterostructures.** Main peaks that are characteristic of molecular vibrations are indicated by arrows.

## Supplemental Notes

### Supplemental Note 1 – Indirect exchange estimation

Effective RKKY coupling between two metallo-molecular spin sites, *i* and *j* could be written as

$$H_{RKKY} = -J_{eff}(r)(\boldsymbol{S}_i \cdot \boldsymbol{S}_j)$$

The quantity $J_{eff}(r)$ in 2D has been demonstrated by Fischer and Klein [3] in the limit $k_F r \sim 1$ as

$$J_{eff} = \frac{{J_0}^2}{8\pi} k_F^2 [\mathcal{J}_0(k_F r_n)\mathcal{N}_0(k_F r_n) + \mathcal{J}_1(k_F r_n)\mathcal{N}_1(k_F r_n)] \qquad (1),$$

where $\mathcal{J}_x$ is a Bessel function of first kind of order $x$, $\mathcal{N}_x$ is a Bessel function of the second kind of order $x$, $k_F$ is the Fermi wavevector of the RKKY mediating surface states, $r_n$ is the distance between spin sites with $n = 1$ corresponding to nearest neighbour and $n = 2$ corresponding to next nearest neighbour with the constant $J_0$ being the enhanced ferromagnetic exchange coupling at π-*d* orbital hybridisation sites [4]. It has been demonstrated that Rashba spin split surface states Pt(111) have $k_F$ values of $1.4\ \mathrm{nm}^{-1}$ and $1.9\ \mathrm{nm}^{-1}$ around the Fermi level [5]. At graphene/Pt(111) interfaces, it has been demonstrated that the spin polarised surface states hybridising with π orbitals are the ones with $k_F$ value of $1.4\ \mathrm{nm}^{-1}$ [6]. Hence taking $k_F$ to be $1.4\ \mathrm{nm}^{-1}$ and taking the nearest neighbour ($r_1$) and next nearest neighbour ($r_2$) spin lattice distances as 0.59 nm and 0.83 nm respectively, one obtains the nearest neighbour and next nearest neighbour antiferromagnetic exchange coupling $J_1$ and $J_2$ as $-0.81\, J_0$ and $-0.34\, J_0$ respectively, leading sizeable frustration in the antiferromagnetic ground state.

### Supplemental Note 2 – Bloch-Grüneisen Formula

Temperature dependence was resistivity was fit using the well-known Bloch-Gruneisen formula

$$\rho(T) = A\left(\frac{T}{T_D}\right)^n \int_0^{\frac{T_D}{T}} \frac{t^n}{(e^t - 1)(1 - e^{-t})} dt \qquad (2).$$

Where $T$ is temperature, $T_D$ is Debye temperature, $A$ is a system dependent constant, and $n$ is a scattering mechanism dependent integer, which is taken to be 5.

### Supplemental Note 3 – Anisotropic magnetoresistance superimposed to (A/O)NE

The anisotropic magnetoresistance is well known to yield $\cos^2(\phi)$ dependence on longitudinal resistance, with a magnetic field dependent coefficient $A_{\mathrm{AMR}}(B_{\mathrm{ext}})$ for the $\cos^2(\phi)$ term, where $\phi$ is the angle between planar magnetic field and applied current shown in phenomenological equation below [7–9]

$$\Delta R = A_{\mathrm{AMR}}(B_{\mathrm{ext}})\cos^2(\phi) + (A_{\mathrm{ANE}}\nabla_z T + A_{\mathrm{ONE}}B_{\mathrm{ext}}\nabla_z T)\sin(\phi)\sin(\theta) \quad (3).$$

The second part of the equation describes anomalous Nernst ($A_{\mathrm{ANE}}$ – Anomalous Nernst constant) and the ordinary Nernst effects ($A_{\mathrm{ONE}}$ – Ordinary Nernst constant) with angle $\theta$ corresponding to angle between z-axis and the magnetic field saturated magnetisation which may be 90 degrees at sufficiently high fields. The ordinary Nernst effect has external magnetic field dependence whereas the anomalous Nernst effect is independent of the magnetic field.